\documentclass[12pt,pdftex]{article}
\usepackage[dvips]{graphicx}
\usepackage{amsmath,amssymb,amsthm}
\usepackage{mathbbol,bbm}
\usepackage[final]{showkeys}
\usepackage{bm}
\usepackage{calc}

\begin{document}

\begin{center}
{\LARGE From the GKLS equation to  the theory of solar and fuel cells }\footnote{Based on the talk given at the 48 Symposium on Mathematical Physics : \emph{Gorini-Kossakowski-Lindblad-Sudarshan Master Equation - 40 Years After}, Toru\'n, June 10-12, 2016}\\[18pt]
R.~Alicki\\[6pt]
Institute of Theoretical Physics and Astrophysics \\
University of Gda\'nsk, Poland \\[6pt]
\end{center}

\begin{abstract}
The mathematically sound theory of quantum open systems, formulated in the 70-ties of XX century and highlighted by the discovery of Gorini-Kossakowski-Lindblad-Sudarshan (GKLS) equation, found a wide range of applications in various branches of physics and chemistry, notably in the field of quantum information and quantum thermodynamics. However, it took 40 years before this formalism has been applied to explain correctly the operation principles of long existing energy transducers like photovoltaic, thermoelectric and fuel cells. This long path is briefly reviewed from the personal author's perspective. Finally, the new, fully quantum  model of chemical engine based on GKLS equation and applicable to fuel cells or replicators is outlined. The model illustrates the difficulty with an entirely quantum operational definition of work, comparable to the Problem of Quantum Measurement.
\end{abstract}

\section{Introduction}

In the \textbf{miraculous year 1976} two independent papers by Lindblad \cite{L} and Gorini--Kossakowski--Sudarshan \cite{GKS} established the general form of the Markovian Master Equation for the density matrix $\rho(t)$ of the open system satisfying complete positivity condition ($\hbar\equiv 1$) 
\begin{equation} 
\frac{d}{dt}{\rho} = -i[H , \rho] + \frac{1}{2}\sum_{j} ([V_j \rho, V_j^{\dagger}]+[V_j ,\rho V_j^{\dagger}])\equiv-i[H , \rho] +\mathcal{L}\rho\, .
\label{GKLS}
\end{equation}
The Hisenberg picture version of GKLS equation, valid for system's time-dependent operator $X(t)$
\begin{equation} 
\frac{d}{dt}{X} = i[H , X] + \frac{1}{2}\sum_{j} ( V_j^{\dagger}[ X , V_j] +[V_j^{\dagger}, X] V_j)\equiv i[H , X] +\mathcal{L}^* X
\label{GKLS_H}
\end{equation}
is also useful.

In \cite{GKS} the systems with finite-dimensional Hilbert spaces were considered, while in \cite{L} infinite dimension was allowed, however the generator of the semigroup (RHS of \eqref{GKLS}) had to be bounded. Notably, the unbounded case remains an open problem till now.
\par
In the same year 1976 the author published his first paper \cite{Alicki:1976} characterizing the class of GKLS equations satisfying the \emph{quantum detailed balance condition} in the form suggested by his MSc supervisor Andrzej Kossakowski. This condition formulated for the Heisenberg picture demands that both the Hamiltonian part $i[H , \cdot]$ and the dissipative one  $\mathcal{L}^*$ commute and form antihermitian and hermitian part of the Heisenberg picture generator, respectively. This generator is treated as a linear bounded (super)operator acting on the Hilbert space consisting of the system operators and equipped with the following scalar product
\begin{equation} 
\langle X , Y\rangle_{\bar{\rho}} \equiv \mathrm{Tr}(\bar{\rho} X^{\dagger} Y) ,
\label{sproduct}
\end{equation}
where the density matrix $\bar{\rho} > 0$ is, automatically, a stationary state for the dynamics given by \eqref{GKLS}. It means that the generator is a normal operator with negatively defined hermitian part and suitable set of eigenvectors and eigenvalues. 

One could also notice that for the case of Gibbs stationary state, i.e.  $\bar{\rho} = Z^{-1} \exp\{-\beta H\}$ the structure of the GKLS equation satisfying detailed balance coincided with the Markovian Master Equation derived by Davies \cite{Davies:1974} using weak coupling limit procedure for a quantum system interacting with a fermionic heat bath. Very soon, in \cite{KFGV:1977}, it has been shown that there is a universal relation between a quantum system weakly interacting with an infinite bath at the equilibrium (KMS) state and GKLS equation satisfying detailed balance condition.

The final result of \cite{Alicki:1976} was the derivation of the Onsager relations with the corresponding expression for  entropy production, thus providing the link between GKLS equation and thermodynamics.

\section{The birth of quantum thermodynamics} 

Perhaps the first result connecting quantum theory with thermodynamics was the Einstein's derivation of the radiation law, where stimulated emission was introduced as necessary to allow equilibration of matter--radiation system \cite{Einstein}. Much later in \cite{GSS} a 3-level maser was treated as a heat engine coupled to two heat baths at different temperatures. 

The emergence of GKLS eqs  allowed to use another Lindblad result \cite{Lindblad:1975} concerning monotonicity of the relative entropy $S(\rho_1|\rho_2) = \mathrm{Tr}(\rho_1 \ln \rho_1 - \rho_1 \ln \rho_2)$ with respect to completely positive maps. This lead to the expression for positive entropy production
\begin{equation}
\sigma(t) = - \mathrm{Tr} \left[ \mathcal L \rho(t) \left( \ln \rho(t) - \ln \bar{\rho} \right) \right] \geq 0 ~, \mathrm{for} \quad \mathcal{L}\bar{\rho} = 0 ~,
\label{eq:spohn}
\end{equation}
valid also beyond the linear regime of Onsager theory. The formula \eqref{eq:spohn}, proved independently in \cite{Spohn:1978} and  \cite{Adory:1977}, has been used to derive the laws of thermodynamics for quantum system coupled to several heat baths \cite{SpoLeb}, ultimately including slowly varying external control \cite{Alicki:1979}. In the later case the GKLS equation has the following form
\begin{equation}
\frac{d}{dt}\rho(t)  = -i[H(t),\rho(t)] + \mathcal{L}(t)\rho(t),\quad \mathcal{L}(t) = \sum_k \mathcal{L}_k(t) .
\label{master2}
\end{equation}
with the time-dependent Hamiltonian including external control and a family of  dissipative time-dependent GKLS generators $\{\mathcal{L}_k(t)\}$ corresponding to a collection of independent heat baths at the inverse temperatures $\{\beta_k\}$. Each generator  $\mathcal{L}_k(t)$ (e.g. derived using weak coupling limit) kills its own temporary Gibbs state $\bar{\rho}_j(t) = Z_j^{-1}(t) \exp\{-\beta_j H(t)\}$. The following energy balance corresponds to the First Law of Thermodynamics
\begin{equation}
\frac{d}{dt}U(t) = J(t) -  P(t) .
\label{work_heat}
\end{equation}
Here 
\begin{equation}
U(t)= \mathrm{Tr}\Bigl(\rho(t)  H(t)\Bigr)
\label{internal}
\end{equation}
is the internal energy of the system, 
\begin{equation}
P(t) \equiv -\mathrm{Tr}\Bigl(\rho(t) \frac{d H(t)}{dt}\Bigr),\quad 
\label{power}
\end{equation}
is the power  provided by the system to the external work depository, and
\begin{equation}
J(t) \equiv \mathrm{Tr}\Bigl(H(t) \frac{d}{dt}\rho(t)\Bigr) =\sum_k J_k(t) , \quad J_k(t) = \mathrm{Tr}\Bigl(H(t) \mathcal{L}_k(t)\rho(t)\Bigr) .
\label{heat}
\end{equation}
is the sum of net heat currents supplied by the individual heat baths.

The definitions of above combined with the inequality \eqref{eq:spohn} reproduce the Second Law of Thermodynamics in the form ($k_B \equiv 1$)
\begin{equation}
\frac{d}{dt} S(t) - \sum_k \beta_k J_k(t) \geq 0 ,
\label{IIlaw}
\end{equation}
with the thermodynamical entropy identified with the von Neumann entropy $S(t) = - \mathrm{Tr}\left[ \rho(t) \ln\rho(t)\right]$.

In the next 25 years the  interest in thermodynamics of quantum open systems described by GKLS eqs was rather low, with notably exception   of \cite{Kosloff:1984}.  However, in the recent years the active research concerning noise, dissipation and decoherence in controlled quantum open systems has been prompted  by the fast technological progress in construction and precise control of micro or mesoscopic devices for information processing and energy transduction. These  emerging technologies pose  problems of  reliability, scalability and efficiency, related to the fundamental principles of thermodynamics, which have to be properly extended to the quantum domain. Similar questions concern the operation principles of biological machinery, where the theory of quantum open systems has already provided important new insights \cite{Plenio}, \cite{David}. Various theoretical approaches have been developed and many models of quantum engines and quantum refrigerators have been studied. For the review of the recent developments see, e.g. \cite{Kosloff:2013} -- \cite{Goold:2016} and references therein.

\section{A generic model of a quantum engine}

Taking into account the recent developments in quantum thermodynamics, one can  draw the following  picture of a typical quantum engine. The  quantum open system corresponding to ``working medium " interacts weakly with a stationary non-equilibrium environment. The later typically, but not always, consists of two heat baths at different temperatures or a collection of chemical baths with properly chosen chemical potentials. The last element, called work reservoir, work repository or simply ``piston", possesses a single degree of freedom, often represented by a harmonic oscillator. In the ideal situation the piston extracts or supplies energy from the working medium during the whole cycle, with negligible transfer of entropy. Therefore, if the piston is modeled e.g. by a quantum harmonic oscillator, its evolution should follow semi-classical paths with  temporal states well-localized in the oscillator phase-space. This is the reason while the piston can be replaced by the time-dependent Hamiltonian modulation, like in the previous Section (see Section 6 for more rigorous arguments).  If the average net power extracted from the working medium is positive, then the oscillations of the piston are self-sustained (self-oscillations) and the whole system acts as an engine yielding a useful work.
\par
The following class of quantum engine  models has been used to describe the operation principles of photovoltaic, thermoelectric and fuel cells in \cite{AGS:2016}--\cite{Alicki:2016a}. The basic tool is a GKLS eq. of the type \eqref{master2} which can be derived under certain standard assumptions: i) weak system-environment coupling,  ii) ergodic properties of the stationary environment, iii)  slow dynamics of the piston  in comparison to the fast internal dynamics.
\par
Here $H(t) = H_0 + H_{mod}(t)$ where $ H_{mod}(t)$ is the Hamiltonian modulation representing the piston.
If, moreover,  $H_{mod}(t)$ is  ``small", in comparison to $H_0$,  we can assume, according to the standard perturbation theory, that in the lowest order approximation $H_{mod}(t)$ commutes with $H_0$. Then, assuming also harmonic oscillations we can put
\begin{equation}
H_{mod}(t) = \xi(t)M = g(\sin\Omega t)\, M, \quad [H_0 , M] = 0,
\label{driving}
\end{equation}
where $g $ is a ``small" amplitude of oscillations and $M= M^{\dagger}$.
\par
The dissipative generator $\mathcal{L}(t)$ obtained by the weak coupling limit procedure depends on the magnitude of perturbation $\xi$
\begin{equation}
\mathcal{L}(t) \equiv \mathcal{L}[\xi(t)] .
\label{generator}
\end{equation}
The generator $\mathcal{L}[\xi]$ is assumed to posses a unique stationary state $\bar{\rho}[\xi]$ what implies
\begin{equation}
\mathcal{L}[\xi]\bar{\rho}[\xi] = 0,  \quad \mathcal{L}'[\xi]\bar{\rho}[\xi]  =-\mathcal{L}[\xi]\bar{\rho}'[\xi]  
\label{identity}
\end{equation}
where $\mathcal{L}'[\xi] \equiv \frac{d}{d\xi}\mathcal{L}[\xi]$ , $\bar{\rho}'[\xi] \equiv \frac{d}{d\xi}\bar{\rho}[\xi] $.
The stationary average power output obtained by the time averaging of $P(t)$ (see \eqref {power}), and taking the second order approximation with respect to $g$ is given by the following compact expression 

\begin{equation}
\bar{P} = -\frac{1}{2}g^2 \mathrm{Tr}\Bigl(\bar{\rho}'[0] \frac{\Omega^2}{\Omega^2 + {\mathcal{L}^*}^2[0]}{\mathcal{L}^*}[0]M\Bigr) ,
\label{power1}
\end{equation}
where, generally, ${\mathcal{L}^*}[\xi]$ is the Heisenberg picture version of the Schroedinger picture generator $\mathcal{L}[\xi]$ .

If additionally, the modulation frequency $\Omega$ is much higher than the relaxation rate of the observable $M$, we can use the even simpler formula
\begin{equation}\label{power2}
\bar{P} = -\frac{1}{2}g^2 \mathrm{Tr}\Bigl(\bar{\rho}'[0] \mathcal{L}^*[0]M\Bigr) .
\end{equation}
The obtained lowest order formulas for power \eqref{power1}, \eqref{power2} are still consistent with thermodynamics. Namely, assuming that the reservoir is a single thermal equilibrium bath  at the inverse temperature $\beta$ the Gibbs state:
\begin{equation}\label{gibbs1}
\bar{\rho}[\xi] = Z^{-1}[\xi] \exp\{-\beta (H_0 + \xi M)\}
\end{equation}
is stationary for $\mathcal{L}[\xi]$, what implies that the formula \eqref{power1} takes form
\begin{align}\label{Seq}
\bar{P}_{eq} &= \frac{g^2}{2}\beta  \mathrm{Tr}\Bigl(\bar{\rho}[0]M \frac{\Omega^2}{\Omega^2 + ({\mathcal{L}^*}[0])^2}\mathcal{L}^*[0] M\Bigr)\\ 
&= \frac{g^2}{2}\beta \langle M ,\frac{\Omega^2}{\Omega^2 + ({\mathcal{L}^*}[0])^2}\mathcal{L}^*[0] M\rangle_{\bar{\rho}[0]} \leq 0 \, ,
\label{power3}
\end{align}
or the simplified formula \eqref{power2} now reads
\begin{equation}
\bar{P}_{eq} = \frac{g^2}{2}\beta \langle M ,\mathcal{L}^*[0] M\rangle_{\bar{\rho}[0]} \leq 0 \, .
\label{power4}
\end{equation}
The inequalities \eqref{power3}, \eqref{power4} are consequences of the fact that for a single heat bath the generator $\mathcal{L}[\xi]$ satisfies quantum detailed balance condition discussed in the Introduction. It follows that the Heisenberg picture generator ${\mathcal{L}^*}[\xi]$ is a negatively defined operator with respect to the scalar product $\langle\cdot , \cdot\rangle_{\bar{\rho}[\xi]}$. This is consistent with the Second Law formulation: ``one cannot extract work from a single heat bath in a cyclic process".

\section{How does a photovoltaic cell work? }

The main conclusion which follows from the results concerning quantum models of engines is a necessary presence of thermodynamical cycles 
executed by the interaction of a working medium with an oscillating piston. Similar conclusions have been independently drawn from a purely macroscopic, classical analysis of various types of energy transducers in a review article on self-oscillations \cite{Jenkins}.  On the other hand, the standard theory of photovoltaic, thermoelectric or fuel cells is based on the ``direct transformation" of light, heat or chemical energy into electric current (DC) identified with the work output. 
\par
Fig.1, taken from the Ref.\cite{AGJ:2017},  presents the standard textbook explanation of the solar cell operation principle. As already noticed in \cite{Wuerfel},  a DC current cannot at all be driven in a closed circuit by a purely electrical potential difference. Another proposed mechanism based on chemical potential cannot be correct as well, because in the hydrodynamic picture of electron gas the current is driven by the gradient of electrochemical potential and hence in a stationary situation cannot follow a closed path. 

\begin{figure} [t] 
\begin{center}
	\includegraphics[width=0.60 \textwidth]{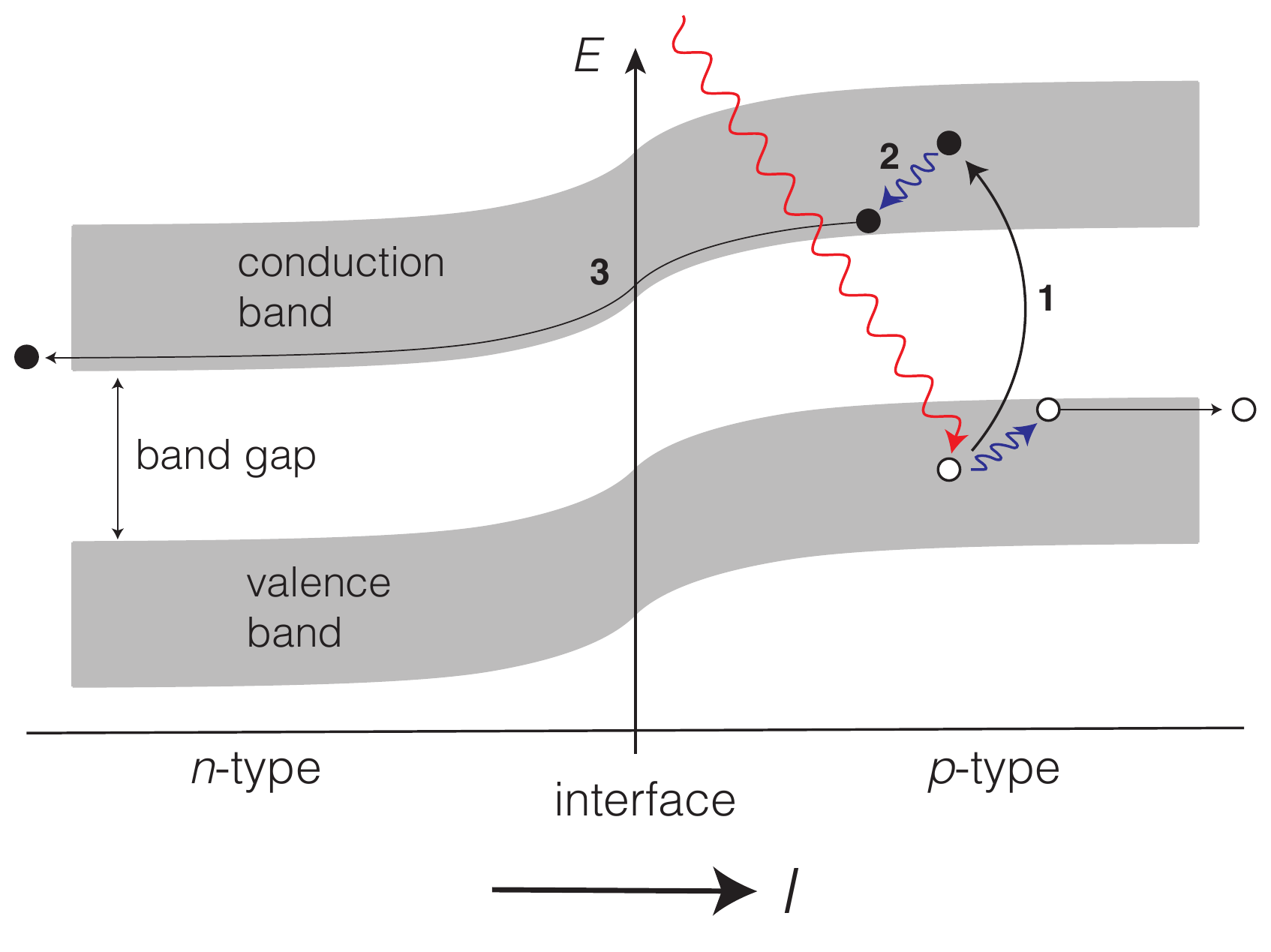}
\end{center}
\caption{\small {\bf 1.}  A solar photon is absorbed in the $p$-type phase of the semiconductor, generating a conducting pair (electron and hole).  {\bf 2.} The pair thermalizes with the phonons in the lattice, dissipating the energy excess above the band gap. {\bf 3.} The electron is driven to the left by the potential difference across the interface, while the hole moves to the right, generating a net voltage between the two terminals.}
\end{figure}

According to the discussion in the previous section work generation in a solar cell requires a self-oscillation (powered by the solar radiation) of the depletion layer at the p-n interface.  This oscillating interface acts as a piston of a heat engine, while the diode effect rectifies the resulting flow of charge.  This combination of a heat-powered self-oscillation with a rectification by the asymmetric interface causes the pumping of the direct current (DC). In \cite{AGJ:2017} the detailed macroscopic and thermodynamical description of this mechanism is presented and illustrated by a simple hydrodynamic analogy that helps to clarify the relevant dynamics.

In the next section the quantum model of a photovoltaic cell derived in \cite{AGS:2016} is briefly discussed.

\section{A photovoltaic cell as a quantum heat engine}

The generic model of a quantum engine presented in Section 3 has been applied to a solar cell in \cite{AGS:2016}.
A working fluid is a free electron gas occupying  conduction (c) and valence (v) bands of the pn-junction. 
The model Hamiltonian can be written in the second quantization formalism using a suitable set of fermionic creation and annihilation operators
\begin{equation}
H_{0} =  \sum_{k} E_{c}(k) c^{\dagger}_{k} c_{k} +  \sum_{l} E_{v}(l) v^{\dagger}_{l} v_{l}. 
\label{ham_electrons}
\end{equation}
A piston is a harmonic oscillator representing collective charge oscillations at the depletion layer of the p-n interface. Such macroscopic oscillation were observed in  experiments as THz plasma oscillations excited by a strong laser pulse \cite{plasma1},\cite{plasma2}. Denoting by $\xi $ the deviation from the equilibrium configuration and replacing the oscillator dynamics by a periodic evolution of $\xi$ one can propose the lowest order interaction Hamiltonian of the form
\begin{equation} 
H_{mod} = {\xi}  \sum_{k}  c^{\dagger}_{k} c_{k} \equiv {\xi} N_c ,
\label{ham_pert1}
\end{equation}
with $\xi(t) = g\sin\Omega t$. The absence of valence band operators is due to the fact that the total number of electrons in the system is preserved and $\sum_{l}  v^{\dagger}_{l} v_{l} $ can be eliminated from the Hamiltonian \eqref{ham_pert1}.

The externally driven electron gas is weakly coupled to cold and hot baths. The first one corresponds to the intraband coupling to phonons at the ambient inverse temperature $\beta$ 
\begin{equation}
H_{intra} =  \sum_{k\vec{k}'}  c^{\dagger}_{k} c_{k'}\otimes R^{(c)} _{kk'}+  \sum_{ll'}  v^{\dagger}_{l} v_{l'}\otimes R^{(v)} _{ll'}, 
\label{ham_intra}
\end{equation}
while the later to the interband coupling to photons 
\begin{equation}
H_{inter} =  \sum_{kl} \bigl( c^{\dagger}_{k} v_{l} +  v^{\dagger}_{l} c_{k}\bigr)\otimes R^{(cv)} _{kl}, 
\label{ham_inter}
\end{equation}
at a certain effective photon inverse temperature $\beta_1 $. Here, $R^{(c)} _{kk'}, R^{(v)} _{ll'}$ and $R^{(cv)} _{kl}$ are suitable bath operators.
\par
The effective temperature $\beta_1 $ is defined by the frequency-dependent ``local" inverse temperature $\beta[\omega]$ related to the photon population number $n(\omega)$ by the Boltzmann factor 
\begin{equation}
e^{-\beta[\omega]{\omega}} = \frac{n (\omega)}{1 + n (\omega)},  
\label{local_temp}
\end{equation}
and computed at the frequency $\omega_g$ corresponding to the energy gap. On Earth, solar photon inverse temperature at standard conditions $\beta_1 = \beta[1eV] \simeq 10^{-3} K^{-1}$.
\par
The GKLS generator obtained using the weak coupling limit procedure and at $\xi = 0$ reads
\begin{equation} 
\mathcal{L}[0]  = \mathcal{L}^{intra}[0] + \mathcal{L}^{inter}[0] ,
\label{L_gen}
\end{equation}
\begin{equation} 
\mathcal{L}^{intra}[0]  = \sum_{\{kk'\}}\mathcal{L}^{(c)}_{kk'}  + \sum_{\{ll'\}}\mathcal{L}^{(v)}_{ll'} ,
\label{L_intra} 
\end{equation}
\begin{equation} 
\mathcal{L}^{inter}[0]  = \sum_{\{kl\}}\mathcal{L}^{(cv)}_{kl}  ,
\end{equation}
\begin{equation}
		\mathcal{L}^{(c)}_{kk'}\rho = \frac{1}{2}\Gamma^{(c)}_{kk'}\Bigl([ c_{k} c_{k'}^{\dagger} \, \rho ,\,c_{k'} c_{k}^{\dagger} ]+ [ c_{k} c_{k'}^{\dagger} , \, \rho \,c_{k'} c_{k}^{\dagger} ]
		\end{equation}
		\begin{equation}
		+ e^{-\beta(E_c(k) -E_c(k')} \bigl([ c_{k}^{\dagger} c_{k'} \, \rho , \,  c_{k'}^{\dagger} c_{k} ] + [ c_{k}^{\dagger} c_{k'}, \, \rho  \,  c_{k'}^{\dagger} c_{k} ] \bigr)\Bigr), 
		\end{equation}
analogical generator for the valence band, and
\begin{equation}
		\mathcal{L}^{(cv)}_{kl}\rho = \frac{1}{2}\gamma_{kl}\Bigl( [c_{k} v_{l}^{\dagger} \, \rho ,\,v_{l} c_{k}^{\dagger}] + [c_{k} v_{l}^{\dagger}, \, \rho \,v_{l} c_{k}^{\dagger}]
\end{equation}
\begin{equation}
+ e^{-\beta_1(E_{c}(k) - E_{v}(l))} \bigl( [c_{k}^{\dagger} v_{l} \, \rho ,\,  v_{l}^{\dagger} c_{k}] + [c_{k}^{\dagger} v_{l} , \, \rho \,  v_{l}^{\dagger} c_{k}] \bigr)\Bigr)\, . 
		\label{ME_inter}
	\end{equation}
Here $\Gamma^{(c)}_{kk'}$ , $\Gamma^{(v)}_{ll'}$ and $\gamma_{kl}$ are relaxation rates for intra and interband processes, respectively.
\par
Although the expression for the average power \eqref{power1} contains only the generator computed for $\xi = 0$ we need the form of the stationary state at the vicinity of this point. This stationary state cannot be computed exactly, but the fact that the intraband phonon-mediated thermalization processes are much faster than the interband radiative recombination suggests, as a sufficient approximation, the following grand canonical ensemble form of the stationary state
\begin{equation}
\bar{\rho}[\xi] = \frac{1}{Z[\xi]} \exp\Bigl\{ -\beta \Bigl[\sum_{k}  \bigl(E_c(k)+ \xi  -\mu_c\bigr) c^{\dagger}_{k}c_{k} + 
\sum_{l} \bigl(E_v(l)-\mu_v\bigr) v^{\dagger}_{l}v_{l}\Bigr]\Bigr\} , 
\label{grand}
\end{equation}
with the ambient (phonon) inverse temperature $\beta$ and two different chemical potentials.
The chemical potentials $\mu_c , \mu_v$ are determined by the average numbers of electrons in both bands and hence by the interband transitions and the  external load. 
\par
The difference of (electro)-chemical potentials between bands can be interpreted as the measured output voltage $V$ 
\begin{equation} 
\mu_c - \mu_v = e V .
\label{voltage}
\end{equation}
Finally, the output power of the solar cell computed using the simplified formula \eqref{power2} reads
\begin{equation}
\bar{P} = {g^2 }{\beta} \langle N_c\rangle_0\,\bar{G}\,\Bigl(\exp\Bigl\{\beta\Bigl(\Bigl[ 1-\frac{\beta_1}{\beta}\Bigr]\omega_g  - eV\Bigl)\Bigr\} -1\Bigr).
\label{powerPV}
\end{equation}
Here $\bar{G} = \sum_{kl} \gamma_{kl} \bigl[ 1 -f_{v}(l)\bigr] f_{c}(k)$, $\langle N_c\rangle_0 = \sum_k f_c(k) $ and  $f_{c(v)}$ are Fermi distributions corresponding to \eqref{grand} at $\xi = 0$..
\par
The condition for positive output power can be written as 
\begin{equation}
eV < eV_{oc} = \omega_g \Bigl( 1-\frac{\beta_1}{\beta}\Bigr) .
\label{open_circuit}
\end{equation}
where $V_{oc}$ is interpreted the open-circuit voltage.  The Carnot factor $1 - \beta_1/\beta$ provides an upper bound on the thermodynamic efficiency of the solar cell. Those results agree with the standard expressions based on the energy balance and the detailed balance condition \cite{Wuerfel}, \cite{SQ-limit}.

\section{A model of chemical engine/replicator}
This last Section is devoted to a new model of a quantum open system which can be called chemical engine, but can also work in the regime of replicator. The basic difference in comparison with the models of previous Sections is that our attention is now concentrated on the dynamics of the piston, represented here by a quantum harmonic oscillator (compare also to the model of fuel cell in \cite{Alicki:2016a} and the quantum piston model in \cite{GAK}). It allows to discuss rigorously the transition from the autonomous models with quantum pistons to those with periodic modulations. Moreover, in this model the working fluid is not explicitly present, but its degrees of freedom are included into chemical baths what simplifies the mathematical analysis.
\par
The model describes the symbolic reaction:
\begin{equation}
A + B \rightleftharpoons C + X.
\label{reaction}
\end{equation}
Here $A, B$ are reactants and  $C$ is the reaction product, all described by quantum equilibrium baths at a common fixed inverse temperature $\beta$ (isothermal conditions). Another ``reaction product"- $X$ represents  excitation of the piston modeled by a quantum harmonic oscillator.

There are interesting instances of $X$ :

a) $X$ is a photon of a single radiation mode generated by a chemical reaction (chemical laser),

b) $X$ is a plasmon with a given frequency representing collective charge oscillations in fuel cells or biological engines, 

c) $X$ is a replicant's molecule produced in a single state with a fixed free energy.

The Hamiltonian of the total system is the following 
\begin{equation}
\hat{H}_{tot}= \omega a^{\dagger} a + H_A + H_B + H_C  + H_{int}\, ,
\label{hamtot}
\end{equation}
where $H_j , j= A,B,C$ are Hamiltonians of chemical baths and the interaction Hamiltonian is given by
\begin{equation}
H_{int}= a\otimes R^{\dagger}  + {a}^{\dagger}\otimes {R}\,  .
\label{hamint}
\end{equation}
Here,  $a^{\dagger}$  creates a single excitation of the harmonic oscillator and $R$ annihilates the molecules $A$ and $B$ and creates the molecule $C$. Obviously, their hermitian adjoints describe the time-reversed processes.

The equilibrium state of the bath is a joint great canonical ensemble for all types of molecules written as
\begin{equation}
{\rho}_R = Z^{-1} \exp\Bigl\{-\beta \sum_{j=A,B,C}\bigl({H}_j - \mu_j {N}_j\bigr)\Bigr\}, \quad [{N}_j , {H}_{j'} ] = 0 ,.
\label{chemeq}
\end{equation}
where the operator ${N}_j$ counts the number of $j$-type molecules and $\mu_j$ is the corresponding chemical potential , $j = A,B,C$.
\par
One should stress that the Hilbert spaces describing chemical baths neither need to have the structure of Fock spaces nor the operator
${R}$ must be a  product  of (bosonic or fermionic) annihilation and creation operators. It is enough to assume the existence of number operators $N_j$ and the validity of the following identity 
\begin{equation}
[\mu_A {N}_A + \mu_B {N}_B + \mu_C {N}_C,  {R}] = (\mu_C - \mu_A -\mu_B) {R} ,
\label{chemcondition}
\end{equation}
which encodes the reaction \eqref{reaction}. One can think about the interaction Hamiltonian \eqref{hamint} as a time-coarse-grained  description of a complicated reversible quantum process with the final effect \eqref{reaction}.
\par
Applying now the standard derivation based on the weak coupling limit one obtains the GKLS equation for the harmonic oscillator density matrix
\begin{equation}
\frac{d}{dt} {\rho}  = -i\omega [{a}^{\dagger}{a}, {\rho}] +\frac{\gamma_\downarrow }{2}\bigl([{a}, {\rho}{a}^{\dagger}] + [{a} {\rho}, {a}^{\dagger}]\bigr)+ \frac{\gamma_\uparrow}{2}\bigl([{a}^{\dagger}, {\rho} {a}] + [{a} {\rho}, {a}^{\dagger}]\bigr).
\label{MEosc}
\end{equation}
This is the well-known Master Equation for linearly damped and pumped quantum oscillator with the standard expressions for damping and pumping rates
illustrating fluctuation-dissipation relations
\begin{equation}
 \gamma_{\downarrow}  =\int_{-\infty}^{\infty}e^{i\omega t}\,\mathrm{Tr}\bigl({\rho}_R {R}(t){R}^ {\dagger}\bigr) dt ,\quad \gamma_{\uparrow}  =\int_{-\infty}^{\infty}e^{i\omega t}\,\mathrm{Tr}\bigl({\rho}_R {R}^{\dagger}(t) {R}\bigr) dt .
\label{relaxation}
\end{equation}
Due to \eqref{chemeq} \eqref{chemcondition} they satisfy the detailed balance condition for chemical baths at isothermal conditions 
\begin{equation}
\frac{\gamma_{\uparrow}}{\gamma_{\downarrow}}   = \exp\{-\beta{\Delta G}\}\,   
\label{dbchem}
\end{equation}
with $\Delta G$ interpreted as the  Gibbs free energy released in the reaction,
\begin{equation}
\Delta G = \omega + \mu_C - \mu_A -\mu_B .
\label{gibbsfree}
\end{equation}
Under the condition 
\begin{equation}
\Delta G < 0 \Rightarrow \gamma_\uparrow > \gamma_\downarrow
\label{amplification}
\end{equation}
the reaction \eqref{reaction} from left to right is spontaneous and the chemical energy is transferred to the oscillator.

The oscillator mean energy 
\begin{equation}
E(t) = \omega \mathrm{Tr} \left({\rho}(t) {a}^{\dagger}{a}\right)
\label{energydf}
\end{equation} 
grows exponentially  
\begin{equation}
E(t) = e^{(\gamma_\uparrow - \gamma_\downarrow)t} W(0) + \bigl[e^{(\gamma_\uparrow - \gamma_\downarrow)t} -1\bigr] \frac{\omega\gamma_\uparrow }{\gamma_\uparrow - \gamma_\downarrow} , 
\label{energy}
\end{equation}
and the average complex amplitude is amplified (compare to the model of ``superradiance" in \cite{AJ:2017})
\begin{equation}
\alpha(t) \equiv  \mathrm{Tr} \left({\rho}(t) {a}\right) =  e^{\frac{1}{2}(\gamma_\uparrow - \gamma_\downarrow)t} e^{-i\omega t}\alpha(0) .
\label{amplitude}
\end{equation}
\subsection{Self-oscillation regime}

Under spontaneous reaction condition \eqref{amplification} the oscillator can serve as a work reservoir. Consider the phase-space picture of the evolution of  initial coherent state $\rho(0) =|\alpha_0\rangle\langle\alpha_0|$, depicted on Fig.2. The evolving state remains Gaussian with exponentially increasing width and centered along the exponentially expanding spiral. Because of gaussianity all properties of the time-dependent state can be easily computed using \eqref{energy}, \eqref{amplitude}.

\begin{figure} [t] 
\begin{center}
	\includegraphics[width=0.60 \textwidth]{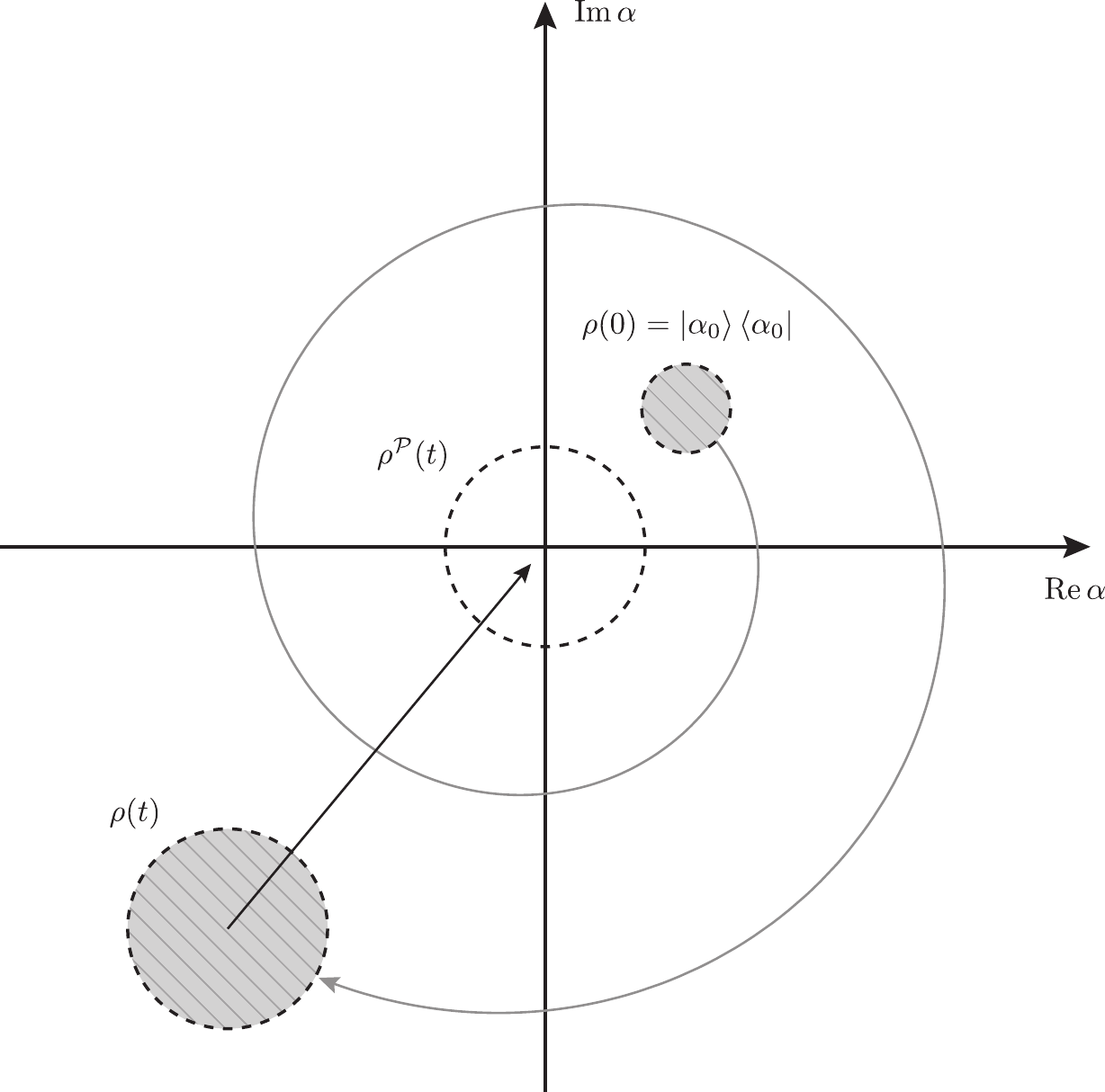}
\end{center}
\caption{\small  \textbf{Phase-space picture of harmonic oscillator evolution in self-oscillating regime.} The initial coherent state $\rho(0) =|\alpha_0\rangle\langle\alpha_0|$ evolves into the mixed Gaussian state $\rho(t)$ which shifted to the origin yields the passive state $\rho^{\mathcal{P}}(t)$.}
\end{figure}

In particular, one can ask the question: \emph{How much work is stored in this system?}

The answer is given by the notion of \emph{ergotropy}  \cite{Alah} and based on the idea of \emph{passivity}, introduced in \cite{Pusz}, in order to characterize equilibrium states of quantum systems. A density matrix  describes a passive state with respect to the Hamiltonian ${H}$ iff it is diagonal in the Hamiltonian basis and its eigenvalues are ordered in the decreasing order with increasing energy. From such states one cannot extract work using cyclical reversible (unitary) operations. Moreover, for any state $\rho$ and any Hamiltonian ${H}$ one can find a passive state unitarily equivalent to ${\rho}$ and denoted by $\rho^{\mathcal{P}}$.
\par
Therefore for a system with a given Hamiltonian ${H}$ the accessible work stored in the state $\rho$ can be identified with the ergotropy
\begin{equation}
W_e =  \mathrm{Tr} ({\rho}{H}) -\mathrm{Tr} ({\rho}^{\mathcal{P}}{H}).
\label{ergotropy}
\end{equation}
It is equal to the maximal energy extractable from the system using cyclical reversible (unitary) operations.
\par
In the case of our example, both the total mean energy and the ergotropy can be computed taking into account \eqref{energy}, \eqref{amplitude} and the fact that $\rho^{\mathcal{P}}(t)$ is obtained by shifting the actual state $\rho(t)$ to the origin of phase-space (Fig.2.).
One can compute also the asymptotic efficiency of work storage
\begin{equation}
\eta_s = \lim_{t\to\infty}\frac{W_e(t)}{E(t)} = \frac{|\alpha_0 |^2}{|\alpha_0 |^2 + \frac{\gamma_\uparrow }{\gamma_\uparrow - \gamma_\downarrow} } \, . 
\label{efficiency}
\end{equation}
The efficiency is close to one if the initial coherent state is semi-classical, i.e. $|\alpha_0|^2 >> 1$. Then, the entropy transport to the oscillator is negligible and hence a classical model with oscillating external perturbation provides a reasonable approximation \cite{GAK}.
\par
The exponential increase of oscillations is, obviously, an idealization which  describes the origin of self-oscillations. In reality, an external load attached to the piston stabilizes the dynamics for long times. An interesting model of such scheme, including measurement, feedback control and external classical perturbation, called quantum flywheel, has been presented in \cite{Levy:2016}.

\subsection{Self-replication regime}

The discussed model of chemical bath driving quantum oscillator describes an engine if one can initialize the piston in a semi-classical state.
If the initial state is Gibbs state than it remains a Gibbs one during the evolution governed by \eqref{MEosc}, i.e.
\begin{equation}
{\rho}(t) = Z^{-1}(t) \exp\Bigl\{-\beta(t)\omega {a}^{\dagger}{a} \Bigr\} ,
\label{gibbschem}
\end{equation}
with the inverse temperature $\beta(t)$ decreasing exponentially. As Gibbs states are passive, no work is stored in the system. On the other hand such evolution is very unstable with respect to an arbitrary shift on the phase-space. To stabilize such passive behavior one can add an additional element to the GKLS generator describing pure decoherence without any impact on the mean energy of the oscillator. The simplest choice reads
\begin{equation}
\mathcal{L}_{dec} \rho = -\Gamma [{a}^{\dagger}{a},[{a}^{\dagger}{a}, {\rho}]] .
\label{puredec}
\end{equation}
Such a term can be derived assuming the environment-induced white-noise perturbations of the oscillator frequency $\omega$.
\par
For $\Gamma > \frac{1}{2}(\gamma_\uparrow - \gamma_\downarrow)$ any initial shift of $\alpha$ decays exponentially to zero. Therefore, instead of the phase-space picture, the classical master equation for the excitation population provides a natural description of this regime. This equation does not contain pure decoherence rate $\Gamma$ and governs the following birth-death process
\begin{equation}
\frac{d}{dt} P_n(t)= \gamma_\downarrow (n +1) P_{n+1}(t) + \gamma_\uparrow n P_{n-1}(t) 
 - \left[ \gamma_\downarrow  n +\gamma_\uparrow (n+1) \right] P_{n}(t)~.
\label{eq:birth}
\end{equation}
Here $P_n(t)$ is the probability of observing $n$ excitations, equal to the suitable diagonal matrix element of $\rho(t)$ in the energy basis. This regime is well suited to the situation where the number operator $a^{\dagger}a$ effectively describes the number of molecules $X$- products of the chemical reaction \eqref{reaction}. The molecule internal degrees of freedom account for the fluctuations of its total energy $\omega$ thus providing the pure decoherence mechanism of \eqref{puredec}. Notice, that the particular form of the interaction Hamiltonian \eqref{hamint} yields the probability rate of creating a new molecule, if $n$ molecules are already present, equal to $\gamma_\uparrow (n+1) $. This means that the birth-death process \eqref{eq:birth} describes self-replication. The discussed model may provide new insight into quantum microscopic mechanisms of biological self-replication processes complementing, e.g. the recent statistical physics approach of \cite{England}.

\section{Concluding remarks}

Almost 40 years after its appearance, GKLS equation  has found application as a mathematical tool in the quantum  theory of solar cells. Similar models, involving self-oscillation mechanism describe thermoelectric generators and fuel cells, further application to biological engines are also to expect. Those theories differ from the standard picture of direct transformation of light, heat or chemical energy into DC and predict new physical effects like the THz radiation emitted by the working cells or, conversely, the resonant stimulation of these devices by external electromagnetic oscillations.
\par
The new model of chemical engine with a quantum piston illustrates the quantum--classical transition and the role of pure decoherence in establishing the operation mode of this system, engine v.s. self-replicator (see also \cite{Kurizki:2017}). This model shows also that the very definition of work in purely quantum terms meets the same conceptual difficulties as the quantum theory of measurement. In the later the observation of the pointer requires another measuring instrument, which in turn requires yet another instrument, and so on, in such a way that the whole process
involves an infinite regression. A practical solution  of the measurement problem introduces a classical instrument at a certain level of description. Similarly, a consistent operational definition of work for quantum machines demands, sooner or later, a classical external modulation.
\par
\textbf{Acknowledgments} The author thanks Dr. Krzysztof Szczygielski for his assistance in preparation of the manuscript and Prof. Gershon Kurizki for comments.

\end{document}